\documentclass[english]{article}
\usepackage[T1]{fontenc}
\usepackage[latin9]{inputenc}
\pdfoutput=1
\usepackage{geometry}
\usepackage{booktabs}
\usepackage{mathrsfs}
\usepackage{amsmath}
\usepackage{graphicx}
\PassOptionsToPackage{normalem}{ulem}
\usepackage{ulem}

\makeatletter

\providecommand{\tabularnewline}{\\}

\@ifundefined{date}{}{\date{}}
\makeatother

\usepackage{babel}
\begin{document}
\title{\textbf{A Constructive GAN-based Approach to Exact Estimate Treatment
Effect without Matching}}
\author{Boyang You\thanks{Ph.D. Student, Department of Economics, University of Bath. Email:
B.You@bath.ac.uk}, Kerry Papps\thanks{Senior Lecturer, Department of Economics, University of Bath. Centre
for Analysis of Social Policy (CASP) Labour, Education and Health
Economics. Email: K.L.Papps@bath.ac.uk}}
\maketitle
\begin{abstract}
Matching has become the mainstream in counterfactual inference, with
which selection bias between sample groups can be significantly eliminated.
However in practice, when estimating average treatment effect on the
treated (ATT) via matching, no matter which method, the trade-off
between estimation accuracy and information loss constantly exist.
Attempting to completely replace the matching process, this paper
proposes the GAN-ATT estimator that integrates generative adversarial
network (GAN) into counterfactual inference framework. Through GAN
machine learning, the probability density functions (PDFs) of samples
in both treatment group and control group can be approximated. By
differentiating conditional PDFs of the two groups with identical
input condition, the conditional average treatment effect (CATE) can
be estimated, and the ensemble average of corresponding CATEs over
all treatment group samples is the estimate of ATT. Utilizing GAN-based
infinite sample augmentations, problems in the case of insufficient
samples or lack of common support domains can be easily solved. Theoretically,
when GAN could perfectly learn the PDFs, our estimators can provide
exact estimate of ATT.

To check the performance of the GAN-ATT estimator, three sets of data
are used for ATT estimations: 1) A linear toy data set with 1-dimensional
input and constant treatment effect is tested. The GAN-ATT estimate
is 0.51\% away from the preset ground truth. 2) A non-linear toy data
set with 2-dimensional input and covariate-dependent treatment effect
is tested. The GAN-ATT estimate is 1.70\% away from the preset ground
truth, which is better than traditional matching approaches including
propensity score matching (PSM) and coarsened exact matching (CEM).
3) A real firm-level data set with high-dimensional input is tested
and the applicability towards real data sets is evaluated by comparing
the other two matching methods. Through the evidences obtained from
the three tests, we believe that the GAN-ATT estimator has significant
advantages over traditional matching methods in estimating ATT.
\end{abstract}

\subparagraph*{{\large{}Key words:}}

Counterfactual inference, Generative adversarial networks, Average
treatment effect on the treated, Machine learning.

\pagebreak{}

\section{Introduction}

\subsection{Aims and Motivations}

Counterfactual inference is a causal inference research method that
constructs hypotheses of no treatment happened in the past, making
the effect that attribute to the treatment available to estimate.
This method is extensively used in the evaluation of medical treatment,
policy performance, and corporate decisions etc. Since counterfactuals
did not really happen, one of the classic approaches for counterfactual
inference is to search for non-treated samples that are as close as
possible to the real samples that are treated, which can be considered
as a matching process. In early studies, matching are usually proceed
based on human experiences, in which selection bias may exist inevitably.
In order to better eliminate selection bias, some advanced matching
methods are further proposed, such as propensity score matching (PSM),
coarsened exact matching (CEM).

In our previous study (You, Papps, 2022), both the two advanced matching
methods are used to estimate the average treatment effect on the treated
(ATT) of a firm-level data set. Evidence shows that the selection
bias of covariates are significantly reduced. However, we also found
that in high-dimensional sample matching processes, due to the existence
of dimensionality reduction, sample dropping, or the lack of common
support domains, the estimation of average treatment effect on the
treated (ATT) may be inaccurate.

To find a solution, this paper proposes a constructive ATT estimator
that integrates classic causal inference framework with a machine
learning approach: generative adversarial networks (GAN), which can
approximate the probability density function (PDF) of sample sets.
By giving identical input condition into the PDFs of both treatment
group and control group, the target conditional average treatment
effect (CATE) can be estimated with no need of matching, and the ensemble
average of the CATEs corresponding to all treated samples is the estimate
of ATT. Using GAN-based infinite sample augmentations, problems caused
by traditional matching can be easily solved. Theoretically, when
GAN could perfectly learn the PDFs, our estimators can provide exact
estimate of ATT.

Furthermore, in order to ensure the effectiveness of this approach,
we constructed two sets of toy data with ATT ground truth to check
the performance of the estimator. Finally, we apply the estimator
to re-estimate the ATT of the previously-used firm-level data set.

\subsection{Research Background and Literature Review}

Generative Adversarial Networks (GAN) is systematically proposed by
Goodfellow (2014), which is a latest deep learning approach involving
dual neural network games. In early stages, GAN is mostly used for
image processing. Through trained GAN models, images with certain
given features can be generated arbitrarily. For example, with a few
strokes on the screen, images of trees, rivers, or beaches can be
randomly generated. With the popularity of GAN, this deep learning
approach has gradually extended to other fields.

One of the innovative uses of GAN is for privacy protection. Huang
et al. (2018) systematically proposed a generative network that can
be used for privacy protection. Liu et al. (2018), Bae et al. (2019),
Ponte (2020) use GAN for privacy control in the data of geolocation,
medical information and marketing. Cai et al. (2021) systematically
summarize all types of generative network used for privacy control
in different perspectives.

While using GAN for privacy protection, scholars find that the function
of GAN that generate identically distributed data sets can be further
used for optimizing casual inference issues. Joint with counterfactual
framework, Yoon \& Van (2018) proposed GANITE to infer individualized
treatment effect (ITE). With this method, Chu et al. (2019), Ge et
al. (2020) estimate ITE based on medical data sets. Ghosh et al. (2021)
use GAN to optimize the performance of propensity score matching.

However, current machine learning based counterfactual studies still
focus on the idea of matching. The popular idea of Yoon \& Van (2018)
can be extended to the estimation of ATT, yet in their work GAN learning
is used for inferring the counterfactual outcome of individual factual
sample, which did not in fact jump out of the matching idea. In order
to supplement the literature in this area, this paper proposes a constructive
ATT estimator based on GAN deep learning approach: GAN-ATT, which
is able to theoretically exact estimate ATT without matching. In details,
GAN-ATT estimator includes four steps in estimation:

1) The two joint PDFs of samples in treatment group and control group
are both learned from two GAN training;

2) Two synthesized data sets with ideally the same PDFs of real sample
sets are generated based on the two trained models;

3) The conditional average treatment effect (CATE) is estimated by
differentiating the two conditional PDFs with identical input condition;

4) ATT is estimated by the ensemble average of CATE over all treatment
group samples.

There are several advantages of this estimator:

First, problems caused by insufficient samples or lack of common support
domains can be completely solved by performing GAN-based sample augmentations,
which can generate infinite samples of the two groups for PDF approximating.
Second, the estimation of ATT can be theoretically proved 100\% accurate
in the ideal case that GAN generates samples with exact the same PDFs
as the real data sets. Third, no matching performed in the estimator
for ATT estimation. So the problem caused by matching, such as the
information loss caused by dimensionality reduction in PSM (Marco
\& Sabine, 2008), or the sample loss caused by high-dimensional common
support domain shortage in CEM (Iacus et al., 2012) can be bypassed.
Fourth, incorporating machine learning approaches, the estimation
results are more objective to the fact, the influence of human factors
are reduced. Last, the privacy of sample providers can be protected
if required.

In addition to proposing the estimator, we further construct two sets
of toy data sets referring to the idea of Yoon \& Van (2018), together
with a real firm-level data set we previously used (You, Papps, 2022)
to check the effectiveness of the GAN-ATT estimator. Evidence proves
that the estimator is feasible and accurate.

\section{Classic Counterfactual Inference Framework}

\subsection{Assumptions to Meet}

To fit our estimator to the casual inference framework, according
to (Rosenbaum and Rubin, 1983; Imbens and Rubin, 2015), the data set
used for estimation should meet two assumptions: unconfoundedness
(or conditional independence assumption) and overlap (or common support
conditions).
\begin{description}
\item [{Assumption}] 1. Unconfoundedness: $Y(d=0,1)\amalg d\mid X$
\end{description}
In the assumption, $d$ is the treatment dummy, $Y(d=0)$ is the outcome
variable of the control group, $Y(d=1)$ is the outcome variable of
the treatment group, $X$ is covariate and $\amalg$ denotes independence.
This assumption can be understand as the difference between $Y(d=1)$
and $Y(d=0)$ under same $X$ should be attribute to the treatment
(Caliendo \& Kopeinig, 2008).\pagebreak{}
\begin{description}
\item [{Assumption}] 2. Overlap: $0<P(d=1\mid X)<1$
\end{description}
This assumption indicates that for samples under any $X$, the probability
of being treated or not being treated should not be equal to zero
(Heckman et al., 1999). Otherwise, there will be a lack of common
support domains for CATE calculations.

Note that the unconfoundedness assumption is relatively strong in
practice. When focusing on estimating ATT only, the two assumptions
can be weaken to 'unconfoundedness for controls' and 'weak overlap'.
See Caliendo \& Kopeinig (2008) for details.

\subsection{Estimation of Average Treatment Effect on the Treated}

Let $\left\{ y_{d},x_{d}\right\} $ be the two sets of sample data
sets that have been observed, where $y_{d}$ are outcome variables
and $x_{d}$ are covariates; $d$ is the treatment dummy, $d=1$ represents
treatment group and $d=0$ represents control group. Let the PDFs
of the covariates $x_{0}$ and $x_{1}$ be $f_{d}\left(x\right)$,
that is:

\begin{equation}
x_{d}\sim f_{d}\left(x\right),\ d=0,1
\end{equation}

Assumes that with given covariate $x_{d}$, the potential outcome
variable $y_{d}$ conforms to the conditional probability density
functions as follows:

\begin{equation}
y_{d}\mid x_{d}\sim g_{d}\left(y\mid x\right),\:d=0,1,
\end{equation}

where $y\mid x$ means the value of the potential outcome variable
$y$ with given covariate $x$.

Note that when $f_{0}\left(x\right)$ and $f_{1}\left(x\right)$ share
the same probability distribution, or share a common trend, the classic
difference-in-difference (DID) method can be used for ATT estimation.
However in practice, sample data sets are usually collected from observational
experiments directly. The selection bias exists in covariates may
lead to significant distribution differences between $f_{0}\left(x\right)$
and $f_{1}\left(x\right)$. In order to eliminate such selection bias,
matching methods are usually required before ATT estimations.

Next, according to Heckman et al. (1997), ATT can be estimated as
follows:

\begin{equation}
ATT=E_{x_{1}\sim f_{1}(x)}\left\{ E_{y_{1}\sim g_{1}\left(y\mid x\right)}\left\{ y_{1}^{f}\mid x_{1}\right\} -E_{y_{1}\sim g_{1}\left(y\mid x\right)}\left\{ y_{1}^{cf}\mid x_{1}\right\} \right\} \text{,}
\end{equation}

where $y_{1}^{f}$ is the outcome variable of samples in the treatment
group that are in real treated based on the fact, while $y_{1}^{cf}$
is the counterfactual outcome variable of samples in the treatment
group that is assumed not being treated.

Since the $y_{1}^{cf}$ cannot be observed, to estimate the treatment
effect, it is necessary to find samples with ideally the same features
in the control group. That is to say, assuming no unobserved confounding
covariates exist, $y_{0}^{f}\mid x$ and $y_{1}^{cf}\mid x$ should
have very similar statistical characteristics under same covairate
$x$ (Heckman et al., 1997). This indicates that in the control group,
$y_{0}^{f}\mid x_{1}$ with given covariate $x_{1}$ can be used as
an estimate of $y_{1}^{cf}\mid x_{1}$. In this case Equation (3)
can be rewritten as:

\begin{equation}
ATT\cong E_{x_{1}\sim f_{1}(x)}\left\{ E_{y_{1}\sim g_{1}\left(y\mid x\right)}\left\{ y_{1}^{f}\mid x_{1}\right\} -E_{y_{0}\sim g_{0}\left(y\mid x\right)}\left\{ y_{0}^{f}\mid x_{1}\right\} \right\} 
\end{equation}

Note that in Equation (4), $E_{y_{1}\sim g_{1}\left(y\mid x\right)}\left\{ y_{1}^{f}\mid x_{1}\right\} -E_{y_{0}\sim g_{0}\left(y\mid x\right)}\left\{ y_{0}^{f}\mid x_{1}\right\} $
can be understand as the conditional average treatment effect (CATE)
under condition $x_{1}$. Besides, the presumption of using Equation
(4) is that the range of $x_{1}$ values is a subset of the range
of $x_{0}$ values, otherwise it cannot be guaranteed that any condition
$x_{1}$ can be found in the control group for $y_{0}^{f}\mid x_{1}$,
which means Asumption 2: Overlap cannot hold.

Classic matching-based ATT estimator such as PSM-DID and CEM-DID are
all based on the idea of estimating $y_{1}^{cf}\mid x_{1}$ via $y_{0}^{f}\mid x_{1}$.
However, no matter which matching method is used, samples without
common support domains and samples without matched pairs will be dropped,
leading to an inevitable trade-off between estimation accuracy and
information loss: the less samples are dropped, the greater the selection
bias; the more samples are dropped, the more data information is lost.
So, it is impossible for matching approaches to accurately estimate
$y_{1}^{cf}\mid x_{1}$ without information loss. To bypass this unsolvable
problem, in the following section we propose a GAN-ATT estimator,
which is able to estimate CATE in Equation (4) via GAN machine learning
so that no matching process is required. For notation simplicity,
in the following sections $y_{0}^{f}$ and $y_{1}^{f}$ are simplified
to $y_{0}$ and $y_{1}$.

\section{GAN-Counterfactual Inference Framework}

\subsection{Principles of GAN}

Generative Adversarial Network (GAN) is first systematically proposed
by Goodfellow et al. (2014). Generally, the training procedure of
GAN can be summarized as the mutual iteration of two neural networks:
generator \ensuremath{\mathscr{G}} and discriminator \ensuremath{\mathscr{D}}.
First, the generator \ensuremath{\mathscr{G}} generates a set of data
based on a white noise input. Next, with the comparison of the real
data set, the discriminator \ensuremath{\mathscr{D}} determines whether
the synthesized data is real or fake (generated). As long as the value
returned by discriminator \ensuremath{\mathscr{D}} is fake, the generator
\ensuremath{\mathscr{G}} optimizes the synthesized data through a
cost function, making it closer to the real. When the discriminator
\ensuremath{\mathscr{G}} cannot distinguish between real and fake,
it can be considered that the training has converged. Based on this
idea, Mirza \& Osindero (2014) extend GAN into a conditional version,
making inputs with conditions available to be trained. Referring to
their work, conditional GAN will be used in this paper for CATE estimations.
The specific algorithm is shown in the next section.

\subsection{Algorithm of GAN Training}

Figure 1(a) illustrates the detailed GAN training procedure. The two
GAN training structures shown in Figure 1(a) are identical. Each training
is consists of two adversarial neural networks: generator $\mathscr{G}_{d}$
and discriminator $\mathscr{D}_{d},\ d=0,1$, where $d=1$ represents
the GAN training for the treatment group and $d=0$ represents GAN
training for the control group. $\left\{ y_{d}\left(i\right),x_{d}\left(i\right)\right\} \,,i=1,\cdots,N_{d}$
is the real data set that is collected from sample pool. $\left\{ \tilde{y}_{d}(i),\tilde{x}_{d}(i)\right\} \,,i=1,\cdots,\tilde{N}_{d}$
is the synthesized data set that is generated from generator $\mathscr{G}_{d}$.

According to Goodfellow et al. (2014) and Mirza \& Osindero (2014),
generator $\mathscr{G}_{d}$ is a multi-layer feedforward neural network
with adjustable parameters. The input of $\mathscr{G}_{d}$ is an
independent and identically distributed Gaussian variable $z_{d}(i)$
together with a treatment dummy $d$. The output node of $\mathscr{G}_{d}$
is linear. The output is a synthesized data set $\left\{ \tilde{y}_{d}(i),\tilde{x}_{d}(i)\right\} $
that will be next judged by the discriminator $\mathscr{D}_{d}$.

Discriminator $\mathscr{D}_{d}$ is also a multi-layer feedforward
neural network. The input of $\mathscr{D}_{d}$ is $\left\{ y_{d}(i),x_{d}(i)\right\} $
and $\left\{ \tilde{y}_{d}(i),\tilde{x}_{d}(i)\right\} $. The output
node of $\mathscr{D}_{d}$ is a sigmoidal function. The output value
is $D_{d}(i)$, which gives a high score from input that is recognized
from a real data set $\left\{ y_{d}(i),x_{d}(i)\right\} $, or gives
a low score from input that is recognized from a fake (synthesized)
data set $\left\{ \tilde{y}_{d}(i),\tilde{x}_{d}(i)\right\} $.
\begin{center}
\includegraphics[width=15cm]{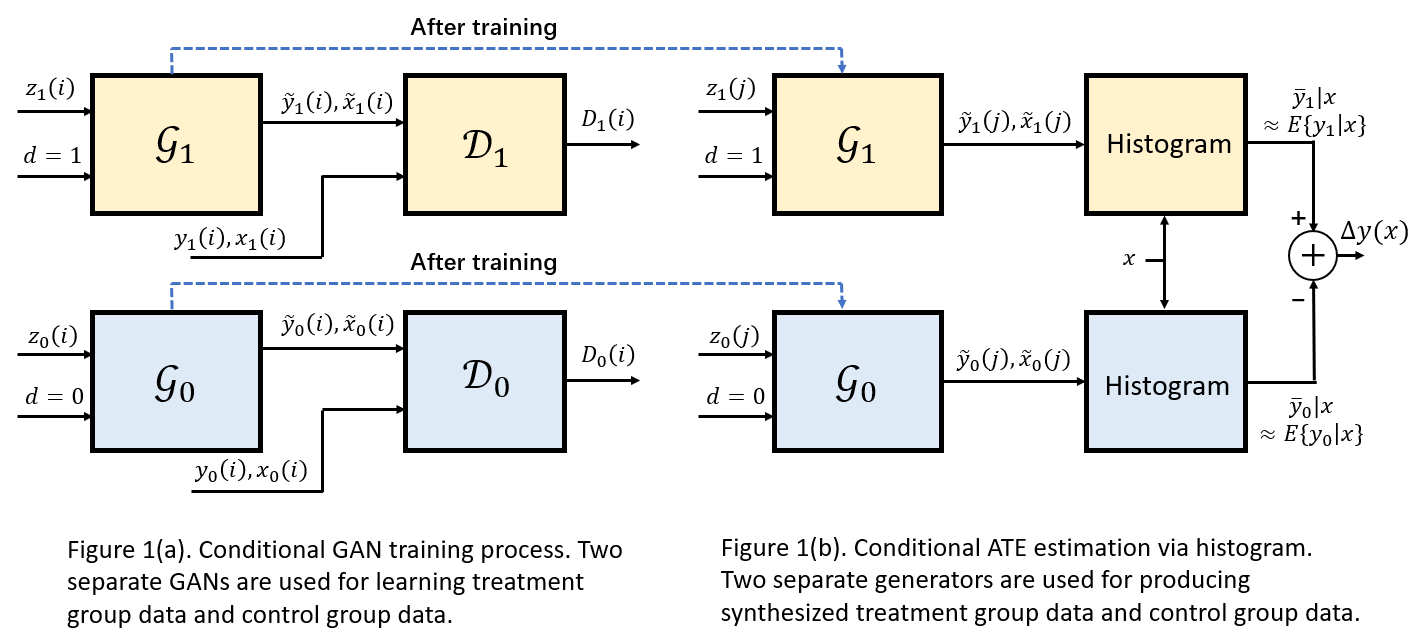}
\par\end{center}

Next, with the output from discriminator $\mathscr{D}_{d}$, generator
$\mathscr{G}_{d}$ reiterate the above procedure through a min-max
game:

\begin{equation}
\min_{\mathscr{D}_{d}}\max_{\mathscr{G}_{d}}V(\text{\ensuremath{\mathscr{G}_{d}}},\text{\ensuremath{\mathscr{D}_{d}}})=\sum_{\left\{ y_{d}(i),x_{d}(i)\right\} \sim g_{d}^{real}\left(y,x\right)}^{i\in\left[1,N_{d}\right]}\log\left[D_{d}(y_{d}(i),x_{d}(i))\right]+\sum_{z_{d}(i)\sim\mathscr{N}\left(z_{d}\right)}^{i\in\left[1,N_{d}\right]}\log\left[1-D_{d}(\tilde{y}_{d}(i),\tilde{x}_{d}(i))\right],
\end{equation}

where $\left\{ y_{d}(i),x_{d}(i)\right\} \sim g_{d}^{real}\left(y,x\right)$
represents that $\left\{ y_{d}(i),x_{d}(i)\right\} $ is collected
from the real data set with the joint PDF of $g_{d}^{real}\left(y,x\right)$;
$z_{d}(i)\sim\text{\ensuremath{\mathscr{N}}}\left(z_{d}\right)$ represents
that $z_{d}(i)$ is collected from the sampling of a Gaussian distribution.
It can be proved that when the real data set is sufficiently large,
the output of the generator \ensuremath{\mathscr{G}} $\left\{ \tilde{y}_{d}(i),\tilde{x}_{d}(i)\right\} \sim g_{d}^{syn}\left(y,x\right)$
will approximate to the real data PDF $g_{d}^{real}\left(y,x\right)$
in terms of the Kullback-Leibler (KL) divergence (Goodfellow et al.,
2014).

To sum up, GAN requires the input of $z_{d}(i)$ and $d$, and the
output generated is an arbitrarily large synthesized data set $\left\{ \tilde{y}_{d}(i),\tilde{x}_{d}(i)\right\} $
that approximately obeys the PDFs of the real data set $g_{d}^{real}\left(y,x\right)$.
The content of Figure 1(b) will be explained in the next section.

\subsection{Combining GAN with Counterfactual Inference}

In order to integrate GAN into counterfactual inference framework,
two separate GAN training procedures for the two data sets $\left\{ y_{d}\left(i\right),x_{d}\left(i\right)\right\} ,\,d=0,1$
are required to perform. After GAN training converges, the synthesized
data sets $\left\{ \tilde{y}_{d}(i),\tilde{x}_{d}(i)\right\} \sim g_{d}^{real}\left(y,x\right)$
can be used for CATE estimations. According to Figure 1, the specific
estimating procedure can divided into 4 steps:

\subsubsection*{1) Joint probability learning}

As it is shown in Figure 1(a), two conditional GANs are first used
to learn the joint PDFs of the treatment group $g_{1}^{syn}\left(y,x\right)\rightarrow g_{1}^{real}\left(y,x\right)$
and the control group $g_{0}^{syn}\left(y,x\right)\rightarrow g_{0}^{real}\left(y,x\right)$
based on their corresponding real data sets $\left\{ y_{1}(i),x_{1}(i)\right\} $
and $\left\{ y_{0}(i),x_{0}(i)\right\} $. Through the training process
mentioned in section 3.2, when reaching convergence, two trained models
which can reflect the corresponding PDFs of treatment group and control
group can be saved for further calculations.

\subsubsection*{2) Synthesized data generating}

As it is shown in Figure 1(b), based on the two obtained trained models.
$\text{\ensuremath{\mathscr{G}}}{}_{1}$ and $\text{\ensuremath{\mathscr{G}}}{}_{0}$
can generate two synthesized data sets $\left\{ \tilde{y}_{d}(i),\tilde{x}_{d}(i)\right\} \sim g_{d}^{syn}\left(y,x\right)\rightarrow g_{d}^{real}\left(y,x\right),\:d=0,1$
. With joint PDFs approximated to the real, the synthesized data sets
follow the conditional average $E\left\{ \tilde{y}_{d}\mid x_{d}\right\} \rightarrow E\left\{ y_{d}\mid x_{d}\right\} $.
Note that in this step, the samples size generated $\tilde{N}_{d}$
can be arbitrarily large. The larger the sample size generated, the
less common support domains between $\left\{ \tilde{y}_{1}(i),\tilde{x}_{1}(i)\right\} $
and $\left\{ \tilde{y}_{0}(i),\tilde{x}_{0}(i)\right\} $ are missing.

\subsubsection*{3) CATE approximating via histogram}

After data sets are generated by $\text{\ensuremath{\mathscr{G}}}{}_{1}$
and $\text{\ensuremath{\mathscr{G}}}{}_{0}$, histogram method is
used on the synthesized sample sets to approximate the conditional
average $E\left\{ y_{d}\mid x_{d}\right\} $. Following the basic
idea of Pearson (1894) and Iacus et al. (2012), the conditional probability
of each covariate will be fitted by a combination of multiple bars
so that a multidimensional distribution can be approximated.

In details, define the dimension of $x_{d}$ in the data set be $q$.
Coarsen $x_{d}$ into multiple small $q$-dimensional equal-sized
cubes. Let $\varDelta\mid x_{d}$ denote such a cube that centered
at $x_{d}$. All samples of $\tilde{y}_{d}\left(i\right)$ with their
covariates $x_{d}\left(i\right)\in\varDelta\mid x_{d}$ will be averaged
to obtain the conditional average $\bar{y}_{d}|x_{d}\rightarrow E\left\{ y_{d}\mid x_{d}\right\} $.
When the number of cubes is sufficient large, the target distribution
can be approximated. Note that the smaller the cube size $\varDelta$
is chosen, the more accurate the approximation of $E\left\{ y_{d}\mid x_{d}\right\} $
can achieve. Yet with the decrease of cube size $\varDelta$, the
sample size of the synthesized data set must be increased to ensure
the existence of the common support domains.

It can be found that the $E\left\{ y_{d}\mid x_{d}\right\} $ estimation
obtained is in fact a function of $x_{d}$. So, define a function
as $\varDelta y(x)=E\left\{ y_{1}\mid x\right\} -E\left\{ y_{0}\mid x\right\} $,
making $\varDelta y(x)$ be a function of conditional average treatment
effect (CATE).

\subsubsection*{4) ATT estimating}

At last, bring $\varDelta y(x)$ into Equation (4), the estimation
of ATT can be obtained by averaging over the treatment group data
set\footnote{In fact, any data set that shares common support domains can be brought
back. If there are no privacy or other concerns, it is an easiest
way to use samples of the treatment group in real data set as input
because they have very close common support domains with the synthesized
data sets.}:

\begin{equation}
ATT=E_{x_{1}\sim g_{1}(y_{1}\mid x_{1})}\left\{ \varDelta y(x_{1})\right\} \approx\frac{1}{N_{1}}\sum_{j=1}^{N_{1}}\varDelta y(x_{1}(j))
\end{equation}

It can be found that by replacing traditional matching procedure,
this estimate is theoretically proved to be able to accurately estimate
ATT provided that the GAN is perfectly trained. In section 4.1, a
standard linear benchmark with ground truth will be taken as an example
to show that the proposed approach can exactly recover the preset
ATT ground truth.

\subsection{Extension to Conditional Tabular GAN}

In practice, it is possible for covariates to be discontinuous, leading
to failures that the generator of traditional GAN cannot deal with.
To make our estimator more applicable to tabular economic data sets,
in this paper we proceed the training through the conditional tabular
generative adversarial network (CTGAN) proposed by Xu et al. (2019).
By introducing conditional generator joint with training-by-sampling
algorithm, CTGAN can effectively identify and learn distributions
of data with both continuous and discrete inputs.

Referring to the Python code\footnote{Detailed code introduction see https://sdv.dev/SDV/user\_guides/single\_table/ctgan.html}
provided by CTGAN (Xu et al., 2019), the joint probability density
functions of the real data sets in this paper are learned through
a method equivalent to the algorithm described in section 3.2. Figure
2 shows the detailed training process. It can be found in the figure
that the only difference from the original method is that two separate
GANs are merged into one and are trained separately by distinguishing
the one-hot input vector $\bar{d}$ ($\bar{d}=$'10' for treatment
group and $\bar{d}=$'01' for control group). The output is same to
the previously-mentioned methods. The major reason of using this equivalent
algorithm is to save computing resources. Detailed algorithm and neural
network structure see Xu et al. (2019, section 4).

With the trained model, any number of synthesized tabular sample with
the joint PDFs same to the real can be generated.
\begin{center}
\includegraphics[width=15cm]{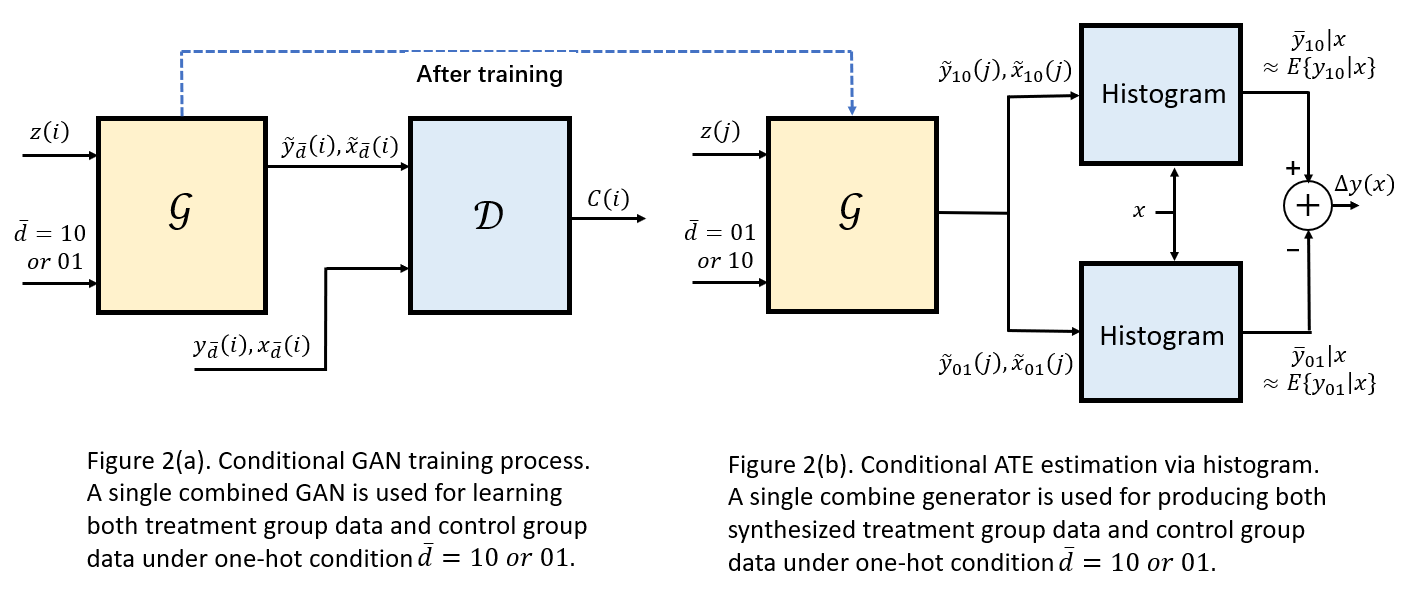}
\par\end{center}

\section{Standard Benchmark Establishment and Estimation Accuracy Tests}

\subsection{Linear Benchmark Establishment}

In order to verify the accuracy of GAN-ATT estimator, a standard linear
benchmark with preset ATT ground truth is established in this section.

First, assume that the data of both treatment group ($d=1$) and control
group ($d=0$) satisfy the following linear model:

\begin{equation}
y_{d}=\alpha+\beta x_{d}+\gamma\cdot d+\varepsilon_{d},\:d=0,1
\end{equation}

In this model, $\alpha$ is the fixed effect, $\beta$ is the parameter
of covariates $x_{d}$, $\gamma$ is the treatment effect that need
to be estimated, $\varepsilon_{d}$ is an Gaussian white noise.

In order to make the benchmark closer to the real, we artificially
set the selection bias as follows:

\begin{equation}
x_{d}\sim\mathscr{N}\left(\mu_{d},\:\sigma_{xd}\right),\:d=0,1,
\end{equation}

which means the covariates $x_{0}$, $x_{1}$ follows two different
Gaussian distributions with mean $\mu_{d}$ and variance $\sigma_{xd}$.

Subsequently, when $\varepsilon_{d}$ is set to be additive white
Gaussian noise, combining with Equation (7), there is:

\begin{equation}
y_{d}\mid x_{d}\sim\text{\ensuremath{\mathscr{N}}}\left(\alpha+\beta x_{d}+\gamma d,\:\sigma_{\varepsilon d}\right),\:d=0,1
\end{equation}

Next, follow the idea in the section 2.2, combine Equation (8)-(9)
with Equation (2), there is:

\begin{equation}
E_{y_{1}\sim g_{1}\left(y\mid x\right)}\left\{ y_{1}^{f}\mid x_{1}\right\} =\alpha+\beta x_{1}+\gamma
\end{equation}

\begin{equation}
E_{y_{0}\sim g_{0}\left(y\mid x\right)}\left\{ y_{0}^{f}\mid x_{1}\right\} =\alpha+\beta x_{1}
\end{equation}

Bring Equation (10) and (11) back in Equation (4), there is $ATT=\gamma$.

The above derivation proves that no matter how much selection bias
exists between $x_{0}$ and $x_{1}$, as long as the left sides of
Equation (10) and Equation (11) can be calculated from sample sets,
the ATT estimated by Equation (4) is theoretically the preset ground
truth $\gamma$. Note that the ATT estimate of Equation (6) is an
approximation of ATT estimate in Equation (4). That is to say, when
the size of sample used for calculation is sufficiently large, the
estimate of GAN-ATT is the preset ATT ground truth $\gamma$.

With the ATT ground truth preset, a linear benchmark is established.
Referring to Equation (7), the model is set with $x_{0}\sim\text{\ensuremath{\mathscr{N}}}\left(0,\:1\right)$,
$x_{1}\sim\text{\ensuremath{\mathscr{N}}}\left(1,\:2\right)$; $\varepsilon_{0},\varepsilon_{1}\sim\text{\ensuremath{\mathscr{N}}}\left(0,\:0.1\right)$;
$\alpha=0$;$\beta=1.5$; $\gamma=1$ is the ATT preset ground truth.

\subsection{Estimator Performance Evaluation}

After the linear benchmark is established, two toy data sets $\left\{ y_{d}(i),x_{d}(i)\right\} $
(50,000 samples in each) are randomly generated from the given model.
In order to avoid confusion, these two data sets will be called 'real'
data sets in this section. Subsequently, two synthesized data sets
(200,000 observations in each\footnote{These samples are randomly generated by the trained model of GAN.
These two synthesized data sets with arbitrarily large sample size
will only be used for intermediate calculations of CATEs and does
not affect the sample size for the final calculation of ATT.}) with ideally the same distribution to the real data set is generated
from GAN machine learning. Next, ATT will be estimated via CATE calculations,
and the result will be compared to the preset ground truth $\gamma=1$.

Figure 3 illustrates the $\varDelta y\left(x\right)$ values (CATE)
in all bar $\varDelta$ in the real data set with common support domain;
Figure 4 illustrates the $\varDelta y\left(x\right)$ values in all
bar $\varDelta$ in the synthesized data set with common support domain.
It can be found that there do exist some differences between the two
figures, especially in the fringes of the bar axis. This is mainly
caused by insufficient samples for CATE calculations on the fringe.
That is, very few synthetic samples fall in these bars, making the
estimation of $\varDelta y\left(x\right)$ biased. However, in the
case of estimaing ATT, this difference has no major effect on the
estimation accuracy. This is because similarly, when bringing the
real treated samples back into $\varDelta y\left(x\right)$, very
few samples will fall into these biased bars, so the biased estimation
will be few. Furthermore, by increasing the sample size of the synthesized
data sets, the bias can be significantly reduced.
\begin{center}
\includegraphics[width=14cm,height=9cm]{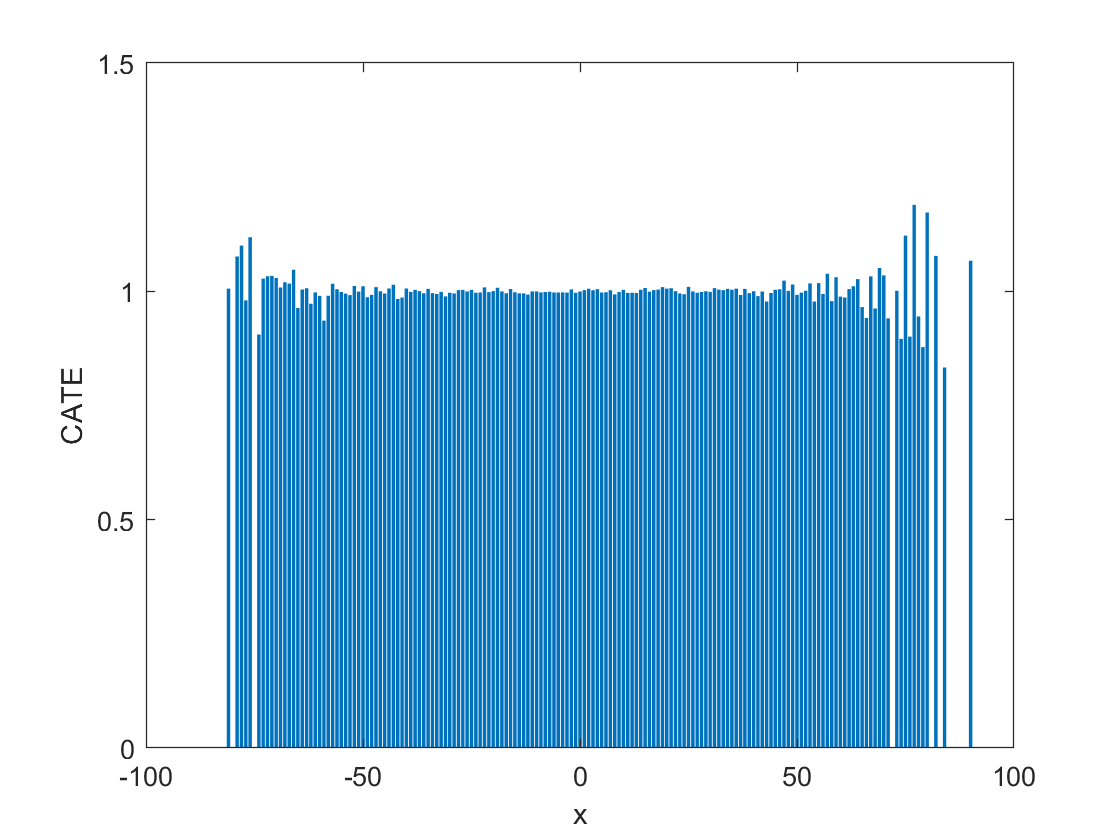}
\par\end{center}

\begin{center}
Figure 3 CATE values of all bars of real data set
\par\end{center}

\begin{center}
\includegraphics[width=14cm,height=9cm]{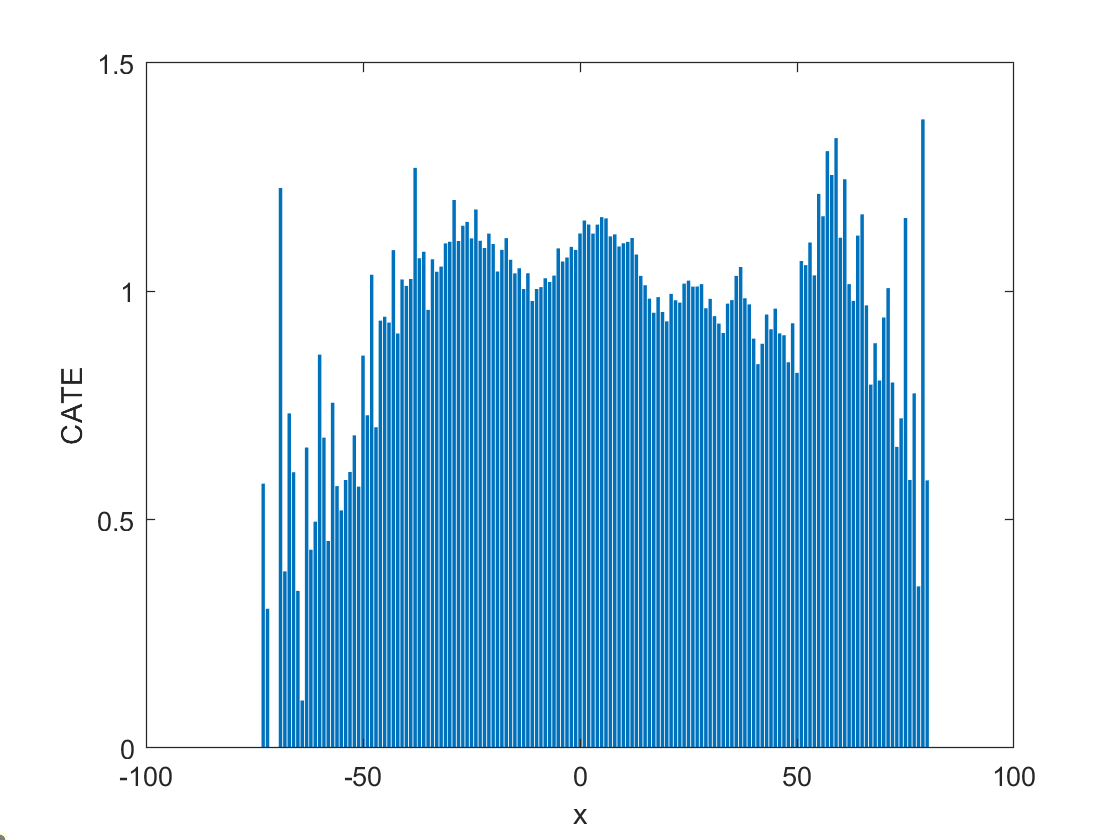}
\par\end{center}

\begin{center}
Figure 4 CATE estimates of all bars of synthesized data set
\par\end{center}

Table 1 is the ATT estimation statistics of this benchmark, including
real data sets and synthesized data sets of both treatment group and
control group. In the last column, the mean of $Estimated\_ATT=0.9949$
is the estimated average treatment effect of the treated (ATT) according
to Equation (6). It can be observed that the estimated ATT is 0.51\%
away from the preset ATT ground truth $\gamma=1$. So, it can be concluded
that the GAN-ATT estimator has a good performance over this toy data
set. Note that the standard error of $Estimated\_ATT$ can also be
given by our estimator. If required, the confidence interval of ATT
estimate can be derived by the obtained mean and variance under Gaussian
distributions.
\begin{center}
Table 1 ATT estimate - Benchmark 1
\par\end{center}

\begin{center}
\begin{tabular}{llllll}
\toprule 
 & {\scriptsize{}$Real\_y_{0}$} & {\scriptsize{}$Real\_y_{1}$} & {\scriptsize{}$Synthetic\_y_{0}$} & {\scriptsize{}$Synthetic\_y_{1}$} & {\scriptsize{}$Estimated\_ATT$}\tabularnewline
\midrule
{\scriptsize{}No. Obs} & {\scriptsize{}50,000} & {\scriptsize{}50,000} & {\scriptsize{}200,000} & {\scriptsize{}200,000} & {\scriptsize{}49,280}\tabularnewline
{\scriptsize{}Mean} & {\scriptsize{}-0.0090} & {\scriptsize{}2.5152} & {\scriptsize{}0.0813} & {\scriptsize{}2.4337} & {\scriptsize{}0.9949}\tabularnewline
{\scriptsize{}Std. Err.} & {\scriptsize{}1.5012} & {\scriptsize{}2.9971} & {\scriptsize{}1.7132} & {\scriptsize{}2.4812} & {\scriptsize{}5.55e-4}\tabularnewline
{\scriptsize{}Inverted Kolmogorov-Smirnov D Statistic} & {\scriptsize{}-} & {\scriptsize{}-} & {\scriptsize{}0.9796} & {\scriptsize{}0.9651} & {\scriptsize{}-}\tabularnewline
{\scriptsize{}Continuous Kullback--Leibler Divergence} & {\scriptsize{}-} & {\scriptsize{}-} & {\scriptsize{}0.6867} & {\scriptsize{}0.6247} & {\scriptsize{}-}\tabularnewline
\bottomrule
\end{tabular}
\par\end{center}

\subsection{Advanced Non-linear Benchmark Establishment}

To further check the performance of the estimator proposed in this
paper, in this section the standard linear benchmark is extended into
a 2-dimensional non-linear data set with the treatment effect dependent
on the covariates.

The benchmark is established as follows:

\[
y_{0}^{k}=\alpha\cdot\frac{1-e^{-\mathbf{W}\cdot\mathbf{X}_{0}}}{1+e^{-\mathbf{W}\cdot\mathbf{X}_{0}}}+\varepsilon_{0}^{k},
\]

\[
y_{1}^{k}=\beta\cdot\frac{1-e^{-\mathbf{W}\cdot\mathbf{X}_{1}}}{1+e^{-\mathbf{W}\cdot\mathbf{X}_{1}}}+t+\varepsilon_{1}^{k},
\]

\begin{equation}
t=\gamma\cdot e^{\frac{-\parallel X_{1}\parallel^{2}}{2\sigma}}
\end{equation}

In this model, $t$ is the preset CATE. $\mathbf{W}$ is a $2\times1$
random vector that satisfies a uniform distribution: $\mathbf{W\mathrm{\sim\mathit{\mathscr{U}}\left(-1,\,1\right)^{2\times1}}}$;
$\boldsymbol{\mathbf{X}}_{0}$ and $\mathbf{X}_{1}$are the vectors
of covariates in control group and treatment group that satisfy$\boldsymbol{\mathbf{X}}_{0}^{k}\sim\text{\ensuremath{\mathscr{N}}}\left(\mu_{0},\:\mathbf{\mathbf{\mathbf{\Sigma}}}\cdot\mathbf{\Sigma}^{T}\right)^{2\times1}$,
$\mathbf{X}_{1}^{k}\sim\text{\ensuremath{\mathscr{N}}}\left(\mu_{1},\:\mathbf{\Sigma}\cdot\mathbf{\Sigma}^{T}\right)^{2\times1},$
$\mathbf{\Sigma\sim\mathrm{\mathscr{U}\left(-1,\:1\right)}}^{2\times2}$,
$\mu_{0}=0^{2\times1}$, $\mu_{1}\sim\text{\ensuremath{\mathscr{U}}}\left(-1,\,1\right)^{2\times1}$;
$\alpha,\beta,\gamma,\sigma$ are parameters to control the value
ranges of $y$ and $t$, here $\alpha=\beta=5,\gamma=1.5,\sigma=4$;
$\varepsilon_{0}$ and $\varepsilon_{1}$ are white noises that satisfy
$\varepsilon_{0}$, $\varepsilon_{1}\sim\text{\ensuremath{\mathscr{N}}}\left(0,\:0.1\right)$.
Note that in order to keep selection bias exist, $\mu_{0}\neq\mu_{1}$.
With these settings, the ATT preset ground truth is 0.6836\footnote{Here, the value of the preset ATT ground truth is approximated by
repeating Monte Carlo simulation since the integral calculation of
$t$ is over-complicated.}. Similarly, it can be proved in the same way in section 4.1 that
the ATT estimated by the GAN-ATT estimator is theoretically equal
to the preset ATT ground truth.

\subsection{Advanced Estimator Performance Evaluation}

After the advanced benchmark is established, two toy data sets $\left\{ y_{d}(i),x_{d}(i)\right\} $
(100,000 samples in each) are randomly generated from the given model.
To avoid losing common support domains, the sample size is also increased
as the covariate dimension increases. Same to section 3.1, two synthesized
data sets (400,000 observations in each) with ideally the PDF same
to the real data set is generated from trained GAN model, and the
ATT is estimated and compared to the preset ground truth. Note that
here since the data dimension is upgraded to two-dimensional, the
bars used for CATE estimations should be upgraded into two-dimensional
cubes $\varDelta$.

In Figure 5, the CATE values in all cube $\varDelta$ over the real
data set with common support domains are presented, while the CATE
values in all cube $\varDelta$ over the synthesized data set with
common support domain are presented in Figure 6.
\begin{center}
\includegraphics[width=14cm]{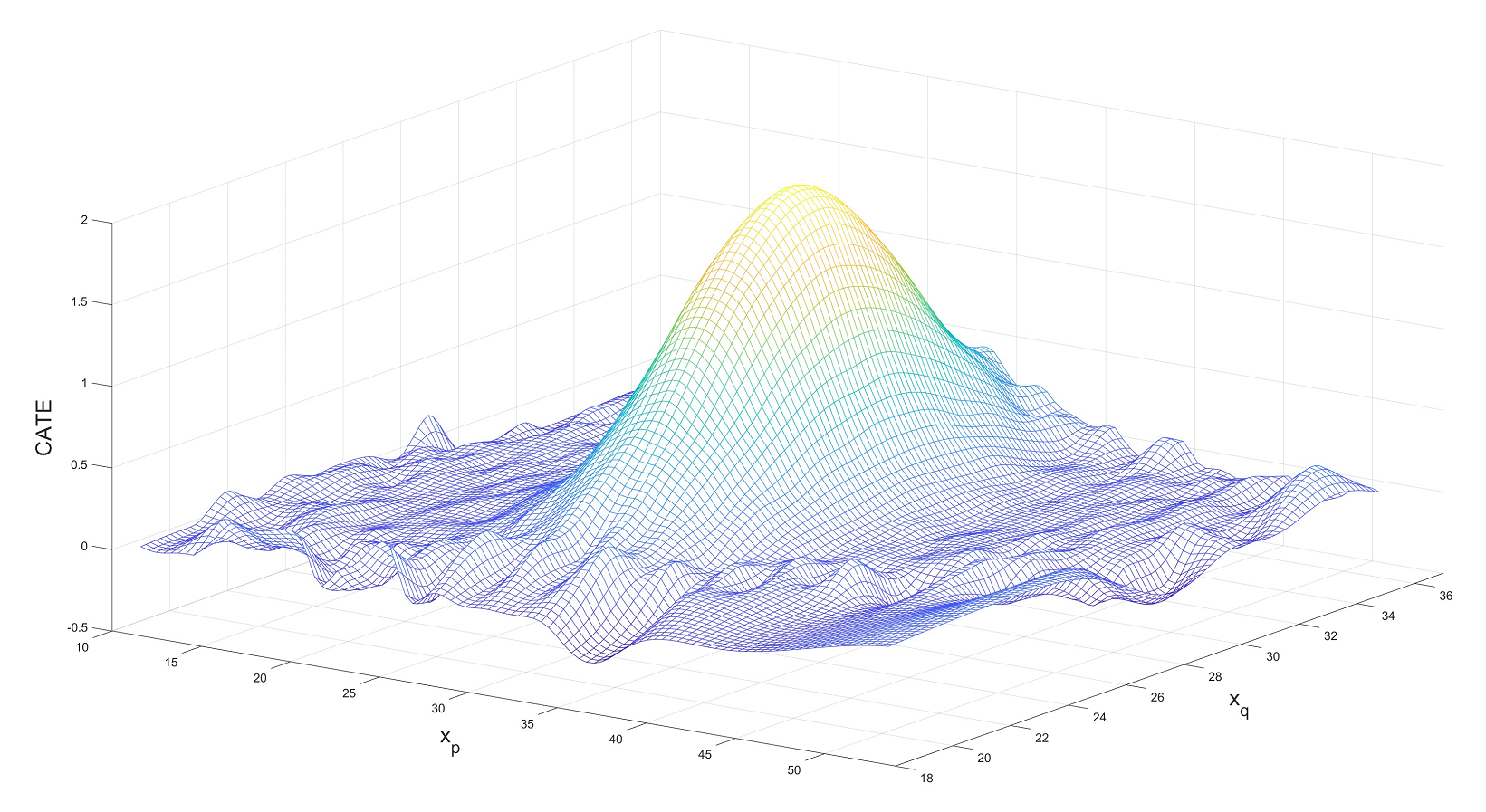}
\par\end{center}

\begin{center}
Figure 5 CATE values of all cubes in real data set
\par\end{center}

\begin{center}
\includegraphics[width=14cm]{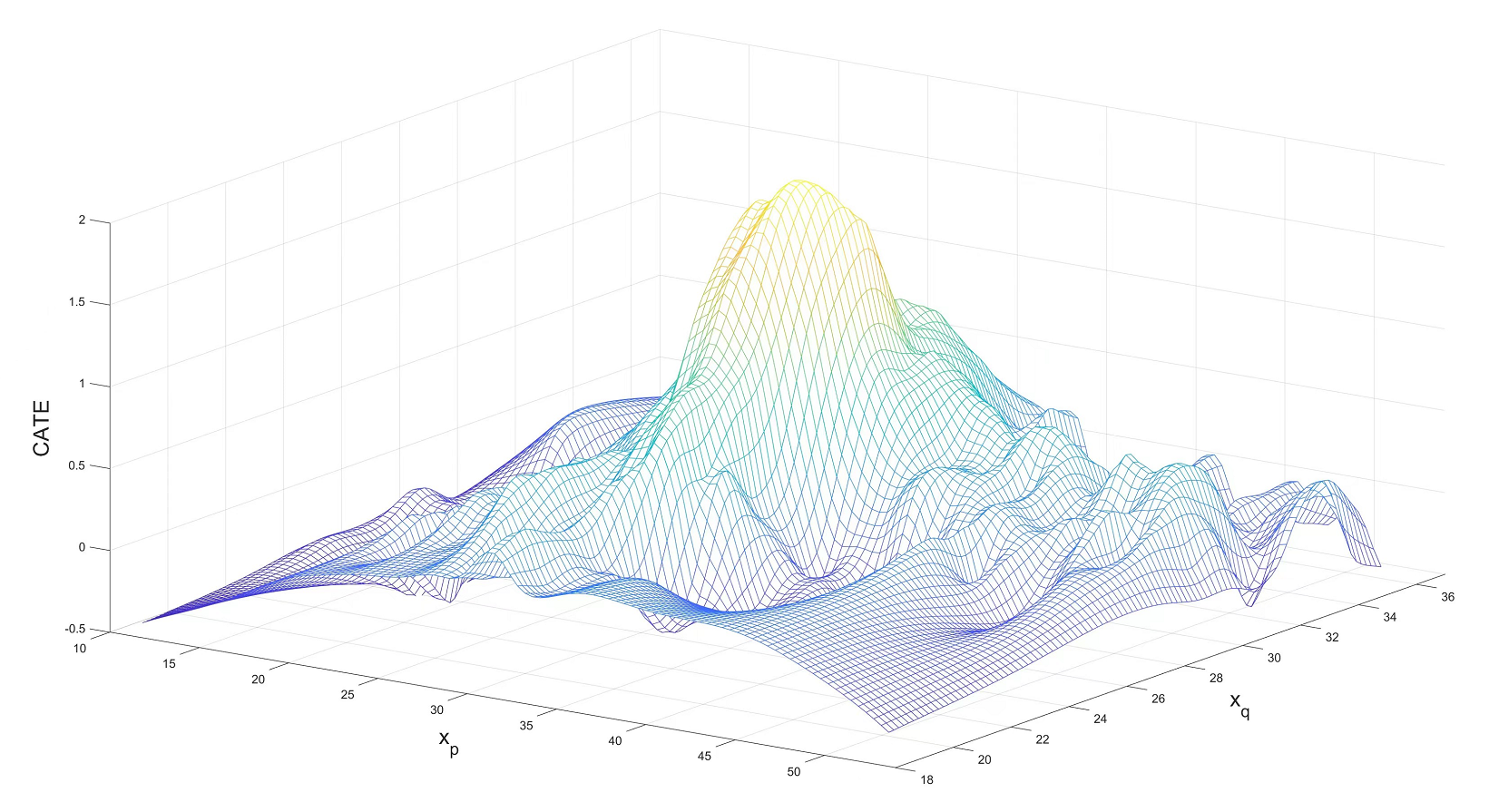}
\par\end{center}

\begin{center}
Figure 6 CATE estimates of all cubes in synthesized data set
\par\end{center}

Table 2 is the ATT estimation statistics of this benchmark. It can
be observed that the mean of $Estimated\_ATT$=0.6720 is the estimated
average treatment effect of the treated (ATT) according to Equation
(6), which is 1.70\% away from the preset ATT ground truth 0.6836.
It can be concluded that the estimator has an acceptable performance
over this toy data set.
\begin{center}
Table 2 ATT estimate - Benchmark 2
\par\end{center}

\begin{center}
\begin{tabular}{llllll}
\toprule 
 & {\scriptsize{}$Real\_y_{0}$} & {\scriptsize{}$Real\_y_{1}$} & {\scriptsize{}$Synthetic\_y_{0}$} & {\scriptsize{}$Synthetic\_y_{1}$} & {\scriptsize{}$Estimated\_ATT$}\tabularnewline
\midrule
{\scriptsize{}Mean} & {\scriptsize{}0.0009} & {\scriptsize{}0.8094} & {\scriptsize{}-0.0173} & {\scriptsize{}0.7905} & {\scriptsize{}0.6720}\tabularnewline
{\scriptsize{}No. Obs} & {\scriptsize{}100,000} & {\scriptsize{}100,000} & {\scriptsize{}400,000} & {\scriptsize{}400,000} & {\scriptsize{}99,166}\tabularnewline
{\scriptsize{}Std. Err.} & {\scriptsize{}2.0300} & {\scriptsize{}2.0472} & {\scriptsize{}2.2268} & {\scriptsize{}2.0317} & {\scriptsize{}0.0021}\tabularnewline
{\scriptsize{}Inverted Kolmogorov-Smirnov D Statistic} & {\scriptsize{}-} & {\scriptsize{}-} & {\scriptsize{}0.9389} & {\scriptsize{}0.9570} & {\scriptsize{}-}\tabularnewline
{\scriptsize{}Continuous Kullback--Leibler Divergence} & {\scriptsize{}-} & {\scriptsize{}-} & {\scriptsize{}0.7483} & {\scriptsize{}0.7139} & {\scriptsize{}-}\tabularnewline
\bottomrule
\end{tabular}
\par\end{center}

Furthermore, in order to make our estimator better adapt to the practical
cases of insufficient samples, we randomly select 1,000 treated samples
in our benchmark 2 as a new treatment group, so without changing the
preset ATT ground truth, the sample size of $Real\_y_{1}$ is reduced
from 100,000 to 1,000.

In the case when the training data is insufficient, problems of overfitting
may exist. To solve problems of this kind, methods of data augmentations,
which add noise to the inputs, are widely used in the field of machine
learning. Referencing to Bishop (1995), samples of the treatment group
are augmented to 100,000 with input white noise. From Table 3 it can
be found that the ATT estimation obtained from our GAN-ATT estimator
is 0.6796, which is 0.59\% away from the preset ground truth. It can
be concluded in the case of insufficient treated samples for machine
learning, GAN-ATT can still provide relatively accurate estimations.
\begin{center}
Table 3 ATT estimates comparing to matching approaches- Benchmark
2
\par\end{center}

\begin{center}
\begin{tabular}{lllll}
\toprule 
 & {\scriptsize{}GAN-ATT} & {\scriptsize{}PSM-NN} & {\scriptsize{}PSM-Kernel} & {\scriptsize{}CEM}\tabularnewline
\midrule
{\scriptsize{}Eistimated\_ATT} & {\scriptsize{}0.6796} & {\scriptsize{}0.7038} & {\scriptsize{}0.7706} & {\scriptsize{}0.7064}\tabularnewline
{\scriptsize{}No. Obs (Treatment Group)} & {\scriptsize{}996} & {\scriptsize{}999} & {\scriptsize{}999} & {\scriptsize{}971}\tabularnewline
{\scriptsize{}No. Obs (Control Group)} & {\scriptsize{}-} & {\scriptsize{}2,918} & {\scriptsize{}99,583} & {\scriptsize{}86,722}\tabularnewline
\bottomrule
\end{tabular}
\par\end{center}

Additionally, for comparison propose, two traditional matching approaches:
PSM and CEM are also used to estimate ATT over this benchmark. From
Table 3 it can be observed that the ATT estimation via PSM-Nearest
Neighbour 1:3 is 0.7038, which is 2.95\% away from the ground truth;
estimation via PSM-Kernel is 0.7706, which is 12.73\% away from the
ground truth; estimation via CEM is 0.7064, which is 3.34\% away from
the ground truth.

From the table it can be observed that comparing to the other three
matching methods, the estimate of GAN-ATT is more accurate. In details,
the ATT estimate of PSM-NN is relatively accurate, yet loses too many
samples in the control group. PSM-Kernel retains nearly full samples
in both two groups, but the estimation is not accurate. This might
be attributed to the inaccurate weightings, because the distance between
propensity scores cannot precisely reflect the real distance before
dimensionality reduction. CEM performs a better trade-off between
information loss and estimation accuracy, yet still cannot get rid
of the problem of sample dropping due to insufficient common support
domains. Through the performance test of the benchmark in this section,
it can be concluded that the GAN-ATT estimator has obvious advantages
over traditional matching approaches.

\subsection{Estimator Performance Evaluation via Firm-level Data}

Referring to our previous study (You \& Papps, 2022), we use the same
data set to further test the performance of our estimator in a more
complex case. The obtained results will also be compared to previous
results estimated by PSM and CEM.
\begin{center}
Table 4 Descriptive statistics - Firm-level data
\par\end{center}

\begin{center}
\begin{tabular}{llllll}
\hline 
 & {\scriptsize{}Profit} & {\scriptsize{}Size} & {\scriptsize{}Age} & {\scriptsize{}Asset} & {\scriptsize{}Output}\tabularnewline
\hline 
{\scriptsize{}No. obs} & {\scriptsize{}1,401,594} & {\scriptsize{}1,411,470} & {\scriptsize{}1,411,472} & {\scriptsize{}1,411,467} & {\scriptsize{}1,411,472}\tabularnewline
{\scriptsize{}Mean} & {\scriptsize{}10,549.64} & {\scriptsize{}260.32} & {\scriptsize{}9.88} & {\scriptsize{}167,026.2} & {\scriptsize{}200,378}\tabularnewline
{\scriptsize{}Std. Err.} & {\scriptsize{}32,685.11} & {\scriptsize{}430.92} & {\scriptsize{}9.25} & {\scriptsize{}2,162,198} & {\scriptsize{}4,208,789}\tabularnewline
\hline 
\end{tabular}
\par\end{center}

Table 4 is the descriptive statistics of the original data set used
in our previous study, which is a panel firm-level data set 2007-2013
extracted from the Chinese Industrial Enterprise Database. According
to the previous discussions, 2007-2009 is the time period before the
treatment, while 2011-2013 is the time period after the treatment.
Samples are previously matched based on firm features of each year
before the treatment to eliminate selection bias. Two ATT estimations
are obtained via difference-in-difference (DID) after matching. The
DID calculation used for ATT estimation is: 1) Calculate the subtractions
of the three-year means of the outcome variables before and after
the treatment; 2) Calculate the subtractions of samples matched between
the treatment group and the control group.

In order to incorporate the GAN-ATT estimator into an equivalent estimation
to the previous, step 1 remains unchanged, while step 2 is replaced
by the calculation of the CATEs. In our previous study, each firm
feature is divided into three covariates for matching purpose (e.g.
firm profit between 2007-2009 is divided into profit2007, profit2008,
profit2009). Firm size, which takes the average of 2011-2013, is selected
as the outcome variable, and the rest four firm features are divided
into covariates with 10 dimensions. Note that firm age is constant
over time, for calculation simplicity, only one dimension is assigned
to firm age. With this, the original data set is transformed into
an equivalent data set with 10-dimensional covariates and 1-dimensional
outcome variable. Besides, since the GAN training requires the integrity
of data input, we eliminate samples with missing input and fill a
small amount of data (less than 4\%) through the mean over time. The
final sample size is 1,250 for the treatment group and 241,289 for
the control group. It can be observed that the sample size of the
treatment group is too small for machine learning. In order to prevent
overfitting, white noise is similarly added to the input according
to Bishop (1995). By this method, the number of input sample for GAN
training is extended into 125,000 (treatment group) and 241,289 (control
group).
\begin{center}
Table 5 ATT estimates comparing to matching approaches - Firm-level
data
\par\end{center}

\begin{center}
\begin{tabular}{lllll}
\toprule 
 & {\scriptsize{}GAN-ATT} & {\scriptsize{}PSM-NN} & {\scriptsize{}PSM-Kernel} & {\scriptsize{}CEM}\tabularnewline
\midrule
{\scriptsize{}Eistimated\_ATT - Firm Size} & {\scriptsize{}6.258} & {\scriptsize{}7.269} & {\scriptsize{}2.029} & {\scriptsize{}10.009}\tabularnewline
{\scriptsize{}No. Obs (}{\scriptsize{}\uline{T}}{\scriptsize{}reatment
Group/}{\scriptsize{}\uline{C}}{\scriptsize{}ontrol Group)} & {\scriptsize{}1,216/-} & {\scriptsize{}1,249/3,746} & {\scriptsize{}1,249/240,956} & {\scriptsize{}1,182/303,699}\tabularnewline
{\scriptsize{}Inverted Kolmogorov-Smirnov D Statistic (T/C)} & {\scriptsize{}0.9795/0.9417} & {\scriptsize{}-} & - & {\scriptsize{}-}\tabularnewline
{\scriptsize{}Continuous Kullback--Leibler Divergence (T/C)} & {\scriptsize{}0.9068/0.9733} & {\scriptsize{}-} & - & {\scriptsize{}-}\tabularnewline
\bottomrule
\end{tabular}
\par\end{center}

Table 5 is the results of ATT estimates via our GAN-ATT estimator
together with PSM (Nearest Neighbour 1:3), PSM (Kernel) and CEM. It
can be observed that the estimation of GAN-ATT estimator is 6.258,
in which CATEs are estimated by 2,000,000 synthesized samples in both
treatment group and control group. It can be found that GAN-ATT still
suffers a loss of 34 treated samples, which is cause by the limited
personal computing power. To improve the results, simply increase
the number of synthesized samples generated by GAN for CATE estimations.
Furthermore, the results obtained previously by the three matching
approaches are also presented for comparison purpose. It can be found
that the GAN-ATT estimate is not far from the estimates of the other
three methods. Short of ground truth, it is hard to say which method
is better in this section, but it can be concluded that our estimator
is competent to estimate ATT in the case of high-dimensional covariate
inputs and insufficient treated samples, while providing theoretically
more accurate estimations with less information loss.

\section{Discussion}

Based on the preliminary performance tests for the GAN-ATT estimator
in section 4, there are several points we need to further discuss:

1) The aim of section 4.1-4.2 is to validate the usability of our
estimator via the simplest benchmarked data set. In this regard, a
standard linear benchmark with one-dimensional covariate and with
selection bias preset is established. By estimating a preset constant
ATT ground truth, the estimation accuracy of our estimator can be
tentatively revealed. Compared to commonly-used semi-benchmarked data
sets, such as TWINS, JOBS, and IHDP (Yao et al., 2021), tests based
on our benchmark can more intuitively reflect the absolute accuracy
of ATT estimates instead of relative accuracy of various estimation
methods, especially methods based on linear regressions. Our evidence
shows that the estimator proposed are within an accuracy loss of 1\%.
So it can be concluded that the GAN-ATT estimator is effective and
accurate in estimating this benchmark.

2) In section 4.3-4.4, to make our estimators more broadly applicable
to economic data sets, the standard linear benchmark is extended into
an advanced non-linear benchmark, which has not been used in the existing
literature. The selection bias is same preset, yet the treatment effect
is set dependent to the covariates with a non-linear relationship.
Furthermore, the dimension of the covariates input is increased to
two. The goal of establishing this benchmark is to test the learning
ability of neural networks towards the non-linearities implicit in
the data set, so as to check the accuracy of estimating conditional
average treatment effect (CATE) and ATT. Our evidence shows that the
accuracy loss of our estimator is 1.70\%. Compared to traditional
matching approaches: propensity score matching (PSM) and coarsened
exact matching (CEM), the GAN-ATT estimate is proved more accurate
while having less information loss. Thus, it can be concluded that
the estimator has a good performance in this benchmark.

3) In section 4.5, we implement our GAN-ATT estimator to a real firm-level
data set, which is previous used in our study (You, Papps, 2022) for
a matching-based counterfactual inference estimation. With proper
data cleaning and augmentation, we find that the estimator is fully
competent for ATT estimation in the case of high-dimensional covariate
inputs and insufficient treated samples. Unfortunately, due to the
need to privacy protections, the database stopped updating after 2013.
However, if our GAN-ATT estimator could be used for casual inference
proposes, privacy of the data providers would be secured. By officially
providing the pre-trained GAN models, the usage of real data set with
privacy concerns can be replaced by a synthesized data set under same
PDF. Joint with the ATT estimation procedure proposed in this paper,
sample privacy issues can be effectively resolved.

4) GAN is a relativity new deep learning technique, especially for
tabular data sets. It must be admitted that there are still shortages
of GAN performance in practice, such as the training validity problem
existing in our estimator. Ideally, the accuracy of our estimator
would approach 100\% if GAN were able to learn the PDF of a data set
with complete accuracy. However in practice, there is currently no
systematic way to verify the performance of GANs in the case with
multidimensional inputs. We found that in the cases when dimension
of covariates is high, performance evaluation criteria provided by
CTGAN (Xu et al., 2019), including KL distance, KS statistic, etc.
may fail. In this paper we try to avoid this failure by repeating
training process with adjusted hyperparameters and layer scales. The
results with closest mean and variance of the real data set are selected
for further calculations. Nevertheless, this is not sufficient to
prove that the two sets of data are equally distributed. We believe
that with the development of GAN training techniques, better testing
methods can be found in our future research.

5) As it is mentioned in section 2.1, to use our estimator, the unconfoundedness
premise must be satisfied. Yet this premise is usually difficult to
check in the field of economics. In order to more effectively eliminate
confounding factors, several methods have been developed in the early
stage, such as learning representation (Bengio et al., 2013), T-Leaner,
R-Learner (Künzel et al. 2019), etc.. To better solve this problem,
we are considering a promising approach to fit all possible covariates
using neural networks. If the R-squared is sufficiently close to 1,
it can be considered that there is no major unobserved covariates.
In order to better use our GAN-ATT estimator, this issue awaits our
further research.

\section{Conclusion}

In this paper we propose the GAN-ATT estimator that can better estimate
average treatment effect on the treated (ATT) without matching. With
the support of the latest mchine learning technique: Generative Adversarial
Network (GAN), the conditional probability density functions (PDFs)
of samples in both treatment group and control group can be obtained.
By differentiating conditional PDFs of the two groups under identical
input condition, the conditional average treatment effect (CATE) can
be estimated, and the ensemble average of corresponding CATE over
treatment group samples is the estimate of ATT. In particular, when
GAN could perfectly learn the conditional PDFs through training, the
ATT estimate can be proved completely accurate.

Furthermore, we generate two benchmarks with preset ATT ground truth.
Together with a real firm-level data set, the performance of GAN-ATT
estimator is directly tested. Evidence shows that: 1) For the two
benchmarks, whether the model is set linear or non-linear, and whether
the treatment effect is set dependent to covariates or constant, the
estimator can closely approximate the ATT ground truths. 2) A real
firm-level data set with high-dimensional covariates and insufficient
number of treated samples is further used for the test. With proper
data augmentations, the estimator is considered competent for estimating
ATT in such case.

Compared to matching approaches, the evidence shows that the GAN-ATT
estimator does have advantages in estimating ATT. Without matching,
there is no compromise between selection bias and information loss
need to take into considerations in our estimator; With GAN-based
sample augmentations, infinite samples can be generated based on the
trained model, so that the problems of insufficient samples or common
support domains are solved; With good GAN learning performance, ATT
estimations can be very close to the fact; With machine learning technique,
the estimates are less influenced by human factors; With providing
per-trained GAN models instead of real samples, fewer privacy concerns
will be taken into consideration.

However, we also found that the performance of GAN deep learning is
away from our expectations. Same to other machine learning approaches,
training results are influenced by the hyper-parameters. Yet in the
case of GAN, with high-dimensional variable inputs, there is no systematic
way so far we know to check training performance. Nevertheless, this
paper has given a theoretical proof that the GAN-ATT estimate is exact
under the premise that GAN performs perfectly. With future advancement
of GAN techniques, we believe our estimator can perform completely
accurate estimate of ATT.

\pagebreak{}

\end{document}